\documentclass{article}
\usepackage{spconf,amsmath,amsfonts,amssymb,float,graphicx,hyperref}

\title{Monaural Continuous-Radius Regional Speech Extraction with
Cross-Radius Consistency Learning}
\name{Biao Dong$^{1}$, Jie Chen$^{1}$, Jianwei Fang$^{2}$, Wei Xiao$^{2}$, Jiqing Han$^{1}$, Yongjun He$^{1}$}
\address{$^{1}$Faculty of Computing, Harbin Institute of Technology, Harbin, China\\
$^{2}$China Unicom, Beijing, China}
\begin{document}
\ninept
\maketitle
\begin{abstract}
Source-to-microphone distance enables speaker-independent speech extraction
without prior enrollment. We propose a monaural continuous-radius regional
speech extraction method that directly models the cumulative speech target
within a queried radius. To realize continuous region control, the query radius
is encoded as a continuous scalar and injected into a time-frequency extraction
network, allowing a single model to operate over arbitrary radii within the
trained range. Exploiting the nested structure of target-speaker sets across 
query radii, we introduce cross-radius consistency learning to stabilize predictions 
for adjacent radii sharing the same nonempty target-speaker set. Experiments on
measured RIRs show that continuous-radius conditioning improves selective
extraction over discrete conditioning. The proposed method achieves 29.25~dB
SI-SDR, 4.40~dB SI-SDRi, 62.01~dB attenuation, and 0.46\% RCE, outperforming
fixed-threshold and local-range baselines. It also remains effective with more
speakers and additive noise.
\end{abstract}

\section{Introduction}
\label{sec:intro}

Target speech extraction (TSE) recovers speech of interest from an acoustic 
mixture according to a user-specified cue. TSE is essential for speech 
communication and listening applications in multi-talker environments. Conventional
TSE often relies on speaker identity~\cite{zmolikova2017speakeraware,wang2019voicefilter},
requiring prior enrollment or target-specific information and limiting
flexibility when the desired speaker is unknown or changes dynamically.
Spatial cues provide an alternative by specifying target speech according to
location rather than identity. In particular, source-to-microphone distance selects 
speech by proximity to the recording device without relying on speaker 
identity~\cite{georganti2011speaker,neri2024speaker}.

Recent studies have shown that source-to-microphone distance enables monaural
speech extraction. Patterson et al.~\cite{patterson2022distance} introduced monaural 
distance-based sound separation (DSS), which separates near and far sources 
relative to a fixed distance threshold.
Lin et al.~\cite{lin2023focus} incorporated distance and speaker
information into monaural TSE. More recently, Shi et al.~\cite{shi2025distance,shi2025room} developed
query-distance-based single-channel target speech extraction (QD-TSE) for local-range extraction, with
room-aware conditioning improving robustness. However, existing monaural
formulations use fixed thresholds or local distance windows rather than a
continuously controlled cumulative region.

Regional speech extraction has also been explored using multichannel spatial
information. Gu and Luo~\cite{gu2024rezero} proposed ReZero, which extracts
sounds from user-specified spatial regions using microphone arrays.
Chen et al.~\cite{chen2024soundbubbles} developed programmable sound
bubbles for hearable devices, preserving sounds within a specified radius while
suppressing those outside. More recent multichannel systems combine directional, distance, and
reverberation cues for regional extraction~\cite{jiang2025regional,lu2026region}.
However, these approaches require multichannel observations, 
leaving continuous-radius monaural regional extraction underexplored.

\begin{figure}[t]
  \centering
  \includegraphics[width=0.7\linewidth]{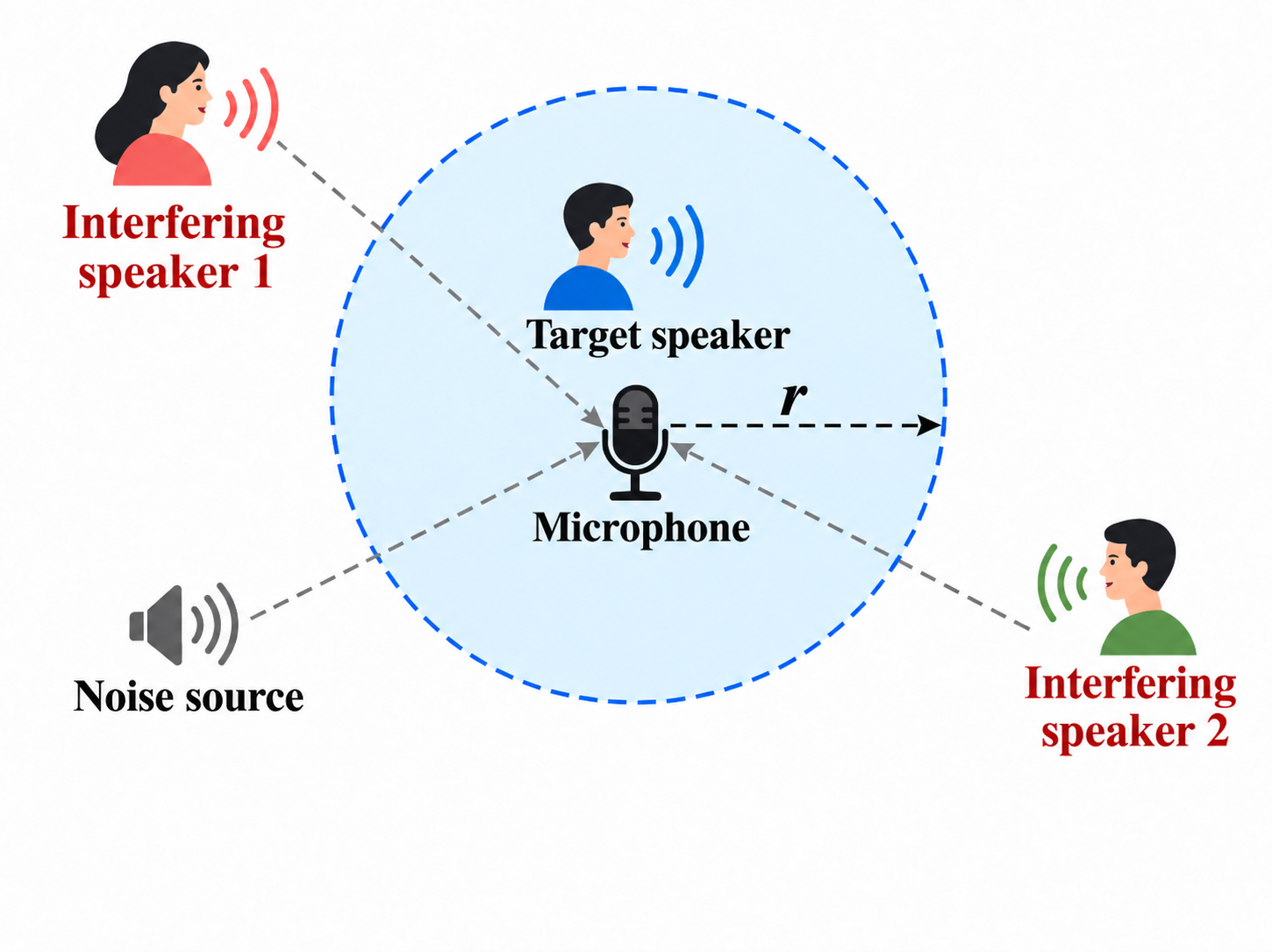}
  \caption{Continuous-radius regional speech extraction task.}
  \label{fig:task_diagram}
\end{figure}

In this work, we investigate monaural continuous-radius regional speech
extraction, which preserves speech within a queried region while suppressing
speech outside it, as illustrated in Fig.~\ref{fig:task_diagram}. The query 
radius defines a spatial region around the microphone, and the task is to 
extract all speakers within this region from the monaural mixture. 
Unlike fixed-threshold DSS and local-range QD-TSE, our approach directly models
cumulative regional targets and supports continuous-radius queries within the
trained range using a single model. Demo samples and code are available at
\url{https://dongb001.github.io/CRadius-RSE/}.

The main contributions of this work are threefold:
\begin{itemize}
    \item We propose a monaural continuous-radius regional speech extraction
    method that directly models cumulative regional targets with continuous
    radius conditioning in a single model.

    \item We exploit the nested structure of target-speaker sets across query 
    radii and introduce cross-radius consistency learning (CRC) to encourage consistent 
    predictions between adjacent radii sharing the same nonempty target-speaker set.

    \item Experiments on measured RIRs show that the proposed method
    outperforms fixed-threshold and local-range baselines across the regional
    extraction tasks, while remaining effective with more speakers and
    additive noise.
\end{itemize}

\section{Method}
\label{sec:method}

\subsection{Problem formulation}
\label{sec:problem_formulation}

Consider a monaural acoustic scene containing $N$ speakers. The observed
mixture is
\begin{equation}
x(t) = \sum_{i=1}^{N} s_i(t) + n(t),
\label{eq:mixture}
\end{equation}
where $x(t)$ denotes the monaural mixture, $s_i(t)$ the microphone-received
signal from the $i$th speaker, and $n(t)$ additive noise. Let $d_i$ denote
the source-to-microphone distance for speaker $i$. 
Given a query radius $r\in[0,R_{\max}]$, where $R_{\max}=4$~m is the maximum
query radius used during training and evaluation, we define the target-speaker
set as
\begin{equation}
\mathcal{S}(r) = \{i \mid d_i \leq r\}.
\label{eq:target_set}
\end{equation}
The desired signal is the sum of speakers within the query radius:
\begin{equation}
y_r(t) = \sum_{i \in \mathcal{S}(r)} s_i(t).
\label{eq:target_signal}
\end{equation}
When $\mathcal{S}(r) = \varnothing$, the target is silent: $y_r(t) = 0$.
The source distances $\{d_i\}_{i=1}^{N}$ are used only to construct
supervision during training and evaluation. At inference, the model receives a
user-specified physical radius $r$ and does not require source-distance
estimation.

Following target speech extraction
formulations~\cite{zmolikova2017speakeraware,wang2019voicefilter,lin2023focus,shi2025distance},
our objective is to learn a model $f_\theta$ that estimates the desired signal
from the mixture $x$ and query radius $r$, 
\begin{equation}
\hat{y}_r(t) = f_\theta(x, r),
\label{eq:estimator}
\end{equation}
where $\hat{y}_r(t)$ denotes the estimated waveform.

The continuous-radius formulation enables a single model to handle continuous
spatial extents within $[0,R_{\max}]$ without restricting queries to a discrete
set of radii. For any
$0 \leq r_1 < r_2 \leq R_{\max}$, the target-speaker sets satisfy
\begin{equation}
\mathcal{S}(r_1) \subseteq \mathcal{S}(r_2),
\label{eq:nested_structure}
\end{equation}
inducing a nested structure across radii. This structure has two key implications. 
First, when the radius increases without crossing any source boundary, the 
target-speaker set remains unchanged:
\begin{equation}
\mathcal{S}(r_1) = \mathcal{S}(r_2).
\end{equation}
Consequently, the desired signals coincide:
\begin{equation}
y_{r_1}(t) = y_{r_2}(t).
\label{eq:local_invariance}
\end{equation}
Second, when the radius increment introduces new speakers, the desired signal 
changes by the sum of the newly included sources. This piecewise-invariant 
behavior motivates the cross-radius consistency learning introduced in
Sec.~\ref{sec:cross_radius_consistency}.

\begin{figure}[t]
  \centering
  \includegraphics[width=0.75\linewidth]{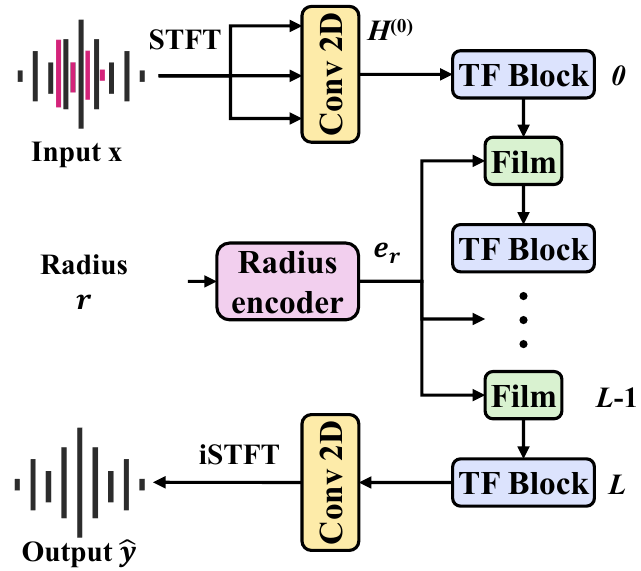}
  \caption{Continuous-radius extraction network.}
  \label{fig:network}
\end{figure}

\begin{figure}[t]
  \centering
  \begin{minipage}[b]{0.47\linewidth}
    \centering
    \includegraphics[width=1.0\linewidth]{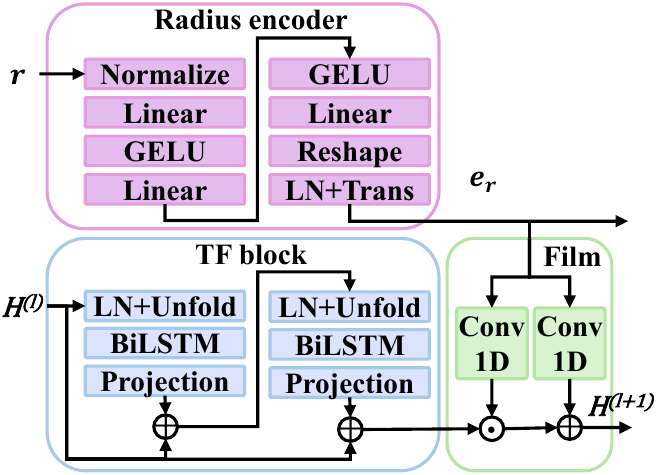}\par\vspace{-2pt}
    {\small (a)}
  \end{minipage}\hfill
  \begin{minipage}[b]{0.47\linewidth}
    \centering
    \includegraphics[width=1.0\linewidth]{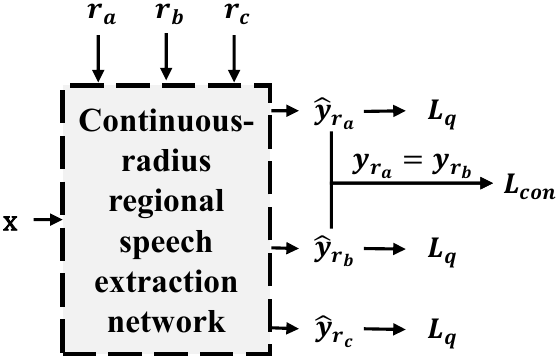}\par\vspace{-2pt}
    {\small (b)}
  \end{minipage}
  \caption{(a) TF block with radius conditioning and (b) cross-radius consistency learning.}
  \label{fig:details}
\end{figure}

\subsection{Extraction network with continuous-radius conditioning}
\label{sec:continuous_radius_network}

The proposed continuous-radius regional speech extraction network is illustrated in 
Fig.~\ref{fig:network}. We employ the lightweight
TF-GridNet~\cite{wang2023tfgridnet} backbone.
Given a mixture waveform $x$, we compute its short-time Fourier transform
(STFT) $X \in \mathbb{C}^{F \times T}$,
where $F$ and $T$ denote frequency bins and time frames. The magnitude-normalized
real, imaginary, and log-magnitude components are concatenated and projected 
into $H^{(0)} \in \mathbb{R}^{C \times T \times F}$, where
$C$ denotes feature channels.

The backbone consists of six sequentially stacked TF blocks indexed by
$l=0,\ldots,L$, where $L=5$. As shown in Fig.~\ref{fig:details}(a), given
the input $H^{(l)}$ to the $l$th block,
bidirectional sequence modeling is first performed along the frequency axis:
\begin{equation}
U^{(l)} = H^{(l)} + P_l^{f}\bigl(\operatorname{BiLSTM}_l^{f}
\bigl(\operatorname{LN}_l^{f}(H^{(l)})\bigr)\bigr), \quad l=0,\ldots,L.
\label{eq:frequency_path}
\end{equation}
A second stage is then performed along the temporal axis:
\begin{equation}
V^{(l)} = U^{(l)} + P_l^{t}\bigl(\operatorname{BiLSTM}_l^{t}
\bigl(\operatorname{LN}_l^{t}(U^{(l)})\bigr)\bigr), \quad l=0,\ldots,L.
\label{eq:temporal_path}
\end{equation}
Here, $\operatorname{LN}(\cdot)$, $\operatorname{BiLSTM}(\cdot)$, and
$P(\cdot)$ denote layer normalization, bidirectional LSTM, and linear
projection, with superscripts $f$ and $t$ indicating frequency and temporal
paths. 

The query radius is normalized and mapped by a three-layer radius encoder
$E_{\phi}$ to a frequency-dependent representation:
\begin{equation}
\tilde{r} = r/R_{\max}, \qquad e_r = E_{\phi}(\tilde{r}).
\label{eq:radius_encoder}
\end{equation}
The representation $e_r$ is injected between consecutive TF blocks through
feature-wise linear modulation (FiLM)~\cite{perez2018film}:
\begin{equation}
H^{(l)} = \Gamma_{l-1}(e_r) \odot V^{(l-1)} + B_{l-1}(e_r), \quad
l = 1, \ldots, L,
\label{eq:film_conditioning}
\end{equation}
where $H^{(0)}$ is the input to block 0, while $H^{(l)}$ for $l\geq1$ is
the FiLM-conditioned input to block $l$. The functions $\Gamma_{l-1}(\cdot)$
and $B_{l-1}(\cdot)$ generate scaling and bias parameters, respectively, and
$\odot$ denotes element-wise multiplication. Thus, the five FiLM modules
condition only the transitions between adjacent TF blocks, with no FiLM before
block 0. The output $V^{(L)}$ of the last TF block is decoded into a complex
spectrum and reconstructed as the waveform $\hat{y}_r(t)$ using an inverse STFT.

\subsection{Cross-radius consistency learning}
\label{sec:cross_radius_consistency}

To explicitly exploit the nested structure across regions, we sample $Q$
query radii $r_1 < r_2 < \cdots < r_Q$ from the continuous interval
$[0, R_{\max}]$ for the same mixture $x$. In our implementation, we set $Q=4$
and sample radii uniformly. Following the local invariance established in
Sec.~\ref{sec:problem_formulation}, consistency is imposed only on adjacent
radius pairs whose target-speaker sets remain unchanged:
\begin{equation}
\mathcal{P} = \{(r_q, r_{q+1}) \mid \mathcal{S}(r_q) = \mathcal{S}(r_{q+1}) \neq \varnothing, \;
1 \leq q < Q\}.
\label{eq:valid_radius_pairs}
\end{equation}
If increasing the radius introduces additional speakers into the region,
the desired signal should change accordingly, and the corresponding pair is
excluded from the consistency constraint to avoid oversmoothing across speaker
boundaries. Computing valid radius pairs $\mathcal{P}$ requires ground-truth
source distances during training, available in simulated data where positions
are known. We further exclude pairs whose common-target RMS does not exceed
$\tau=10^{-4}$.

As illustrated in Fig.~\ref{fig:details}(b), for each valid pair
$(r_a, r_b) \in \mathcal{P}$, we minimize the mean absolute waveform
discrepancy normalized by the RMS amplitude of their common target:
\begin{equation}
\mathcal{L}_{\mathrm{con}} =
\frac{1}{|\mathcal{P}|}
\sum_{(r_a,r_b)\in\mathcal{P}}
\frac{\lVert\hat{y}_{r_a}-\hat{y}_{r_b}\rVert_1}
{N_s\max\!\left(\operatorname{RMS}(y_{r_a}),\epsilon\right)}.
\label{eq:consistency_loss}
\end{equation}
Here, $N_s$ is the waveform length and $|\mathcal{P}|$ is the number of
valid pairs. We set $\mathcal{L}_{\mathrm{con}}=0$ when
$\mathcal{P}=\varnothing$. In implementation, the RMS normalization term is
lower-bounded by $\epsilon=10^{-8}$ for numerical stability. The root-mean-square amplitude
is defined as
\begin{equation}
\operatorname{RMS}(z) = \sqrt{\operatorname{mean}(z^2)}.
\label{eq:rms}
\end{equation}

\subsection{Loss function}
\label{sec:training_objective}

For each query radius $r_q$, the loss depends on whether the target region
contains any speaker. We denote $\hat{y}_q = \hat{y}_{r_q}$, $y_q = y_{r_q}$,
and the prediction error $g_q = \hat{y}_q - y_q$. When the target region
is nonempty, we use a normalized reconstruction loss:
\begin{equation}
\ell_{\mathrm{rec}}(\hat{y}_q, y_q) = 10\log_{10}\left(1 +
\frac{\mathrm{MSE}(g_q)}{\mathrm{MSE}(y_q)+\epsilon}\right),
\label{eq:reconstruction_loss}
\end{equation}
where $\mathrm{MSE}(z)=\operatorname{mean}(z^2)$ and $\epsilon=10^{-8}$. The per-radius loss is then defined as
\begin{equation}
\mathcal{L}_q =
\begin{cases}
\ell_{\mathrm{rec}}(\hat{y}_q, y_q), & \mathcal{S}(r_q) \neq \varnothing,\\[2pt]
\alpha \cdot \operatorname{mean}(|\hat{y}_q|), & \mathcal{S}(r_q) = \varnothing,
\end{cases}
\label{eq:per_radius_loss}
\end{equation}
where $\alpha=100$ is the zero-target suppression weight. For empty targets,
this penalty term encourages the model to output silence. The
extraction loss is obtained by averaging over the $Q$ query radii:
\begin{equation}
\mathcal{L}_{\mathrm{ext}} = \frac{1}{Q}\sum_{q=1}^{Q} \mathcal{L}_q.
\label{eq:extraction_loss}
\end{equation}
The total training loss is
\begin{equation}
\mathcal{L} = \mathcal{L}_{\mathrm{ext}} +
\lambda_{\mathrm{con}} \mathcal{L}_{\mathrm{con}},
\label{eq:total_loss}
\end{equation}
where $\mathcal{L}_{\mathrm{con}}$ is the cross-radius consistency loss from
Eq.~\eqref{eq:consistency_loss}, and $\lambda_{\mathrm{con}}=0.1$.

\section{Experimental Setup}
\label{sec:experimental_setup}

\subsection{Data}
\label{sec:data}

We generate RIRs for 1,000 rooms using 
gpuRIR~\cite{diaz2020gpurir}, with room length and width sampled from
$[2,12]$~m, height from $[2,5]$~m, reverberation time
$\mathrm{RT60} \in [0.1, 1.0]$~s,
and source distances in $[0.3, 5.0]$~m for training and validation. 
Measured RIRs from dEchorate~\cite{dicarlo2021dechorate} are used 
as the test set to evaluate generalization to real acoustic conditions~\cite{gburrek2024diminishing}. 
Speech sources are drawn from AISHELL-1~\cite{bu2017aishell1},
AISHELL-2~\cite{du2018aishell2}, THCHS-30~\cite{wang2015thchs30}, and
VCTK~\cite{yamagishi2019vctk}. 
Training and validation mixtures contain two speakers. 
Additive noise is sampled from the DNS Challenge noise dataset~\cite{dubey2024dns}. 
All signals are sampled at
16~kHz and are 4.5~s long. Source distances are grouped into five intervals to construct diverse source
profiles, while query radii for training and validation are sampled from
$[0,R_{\max}]$. The target is the sum of speakers inside the query
radius. Noise is added with probability 0.5 at an SNR sampled from
$[0,15]$~dB.

\subsection{Baselines}
\label{sec:baselines}

We adopt two monaural distance-based baselines. For DSS~\cite{patterson2022distance}, 
we train four fixed-radius models at 1,
2, 3, and 4~m and use the model corresponding to the requested radius at
inference. Its near output is treated as the regional estimate.
We additionally reproduce QD-TSE~\cite{shi2025distance}, 
which extracts speakers within a local distance 
window $[d_q - r_{\mathrm{spk}}, d_q + r_{\mathrm{spk}}]$ centered at query 
distance $d_q$. Following the original work, we set $r_{\mathrm{spk}} = 0.5$~m. 
To approximate our cumulative regional query at radius $r$, we partition $[0,r]$ 
into non-overlapping intervals of width $\Delta = 1.0$~m and sum the QD-TSE 
outputs from each interval center. This accumulation requires multiple forward 
passes per query, whereas our method processes each continuous-radius query in
a single forward pass. 
All methods are evaluated on the same test mixtures at query radii 1, 2, 3, and 4~m.

\subsection{Evaluation metrics}
\label{sec:evaluation_metrics}
We evaluate three conditions based on the 
target-speaker set induced by the query radius. For full-target cases with 
all sources within the radius, we measure speech preservation using 
SI-SDR~\cite{leroux2019sdr}. For partial-target cases with only some sources 
included, we evaluate selective extraction using SI-SDRi. For empty-target 
cases with no sources in the region, we quantify suppression using attenuation (Atten.).
To measure cross-radius consistency, we evaluate Radius Consistency Error (RCE)
on a 0.1-m query grid over $[0,R_{\max}]$:
\begin{equation}
\operatorname{RCE}(\%) =
\frac{100}{|\mathcal{P}_{\mathrm{eval}}|}
\sum_{(r_a,r_b)\in\mathcal{P}_{\mathrm{eval}}}
\frac{\lVert\hat{y}_{r_a}-\hat{y}_{r_b}\rVert_1}
{N_s\max\!\left(\operatorname{RMS}(x),\epsilon\right)}.
\label{eq:rce}
\end{equation}
where $\mathcal{P}_{\mathrm{eval}}$ contains adjacent query pairs with the same
nonempty target-speaker set, and $N_s$ is the signal length. We report the
median RCE over utterances with $|\mathcal{P}_{\mathrm{eval}}|>0$.

\subsection{Implementation details}
\label{sec:implementation_details}

The FFT length and hop size of the STFT are set
to 192 and 128 samples, respectively. The model uses six TF blocks ($L=5$), a
backbone feature dimension of 32, and a bidirectional LSTM hidden dimension of
64, resulting in approximately 0.66~M trainable parameters. We train the model
with the Adam optimizer using an initial learning rate of $6 \times 10^{-4}$, a
warm-up of 2,000 optimization steps, and a ReduceLROnPlateau scheduler. The
batch size is 1, and the maximum number of training epochs is 50. 
Results are averaged over three independent runs with different random seeds.
\section{Results}
\label{sec:results}

\subsection{Comparison with baselines}
\label{sec:comparison_with_baselines}

We assess whether direct cumulative-region modeling improves extraction over
task-matched fixed-threshold and local-range baselines on measured RIRs
(Table~\ref{tab:baseline_comparison}).
The proposed method achieves the highest scores on all three regional extraction metrics. 
Compared with DSS, it improves full-target SI-SDR by 7.22~dB, 
partial-target SI-SDRi by 4.01~dB, and empty-target attenuation by 5.89~dB. 
Compared with QD-TSE with interval summation, it yields gains of 3.26~dB SI-SDR, 
3.28~dB SI-SDRi, and 7.91~dB attenuation. In addition, the proposed model produces 
each regional estimate with a single forward pass, whereas the interval-summation 
baseline requires multiple local-range queries.

\begin{table}[H]
  \caption{Comparison with distance-based baselines.}
  \label{tab:baseline_comparison}
  \centering
  \small
  \setlength{\tabcolsep}{3pt}
  \begin{tabular}{lc|ccc}
    \hline
    Method &
    \shortstack{Radius\\mode} &
    \shortstack{Full\\SI-SDR\\(dB) $\uparrow$} &
    \shortstack{Partial\\SI-SDRi\\(dB) $\uparrow$} &
    \shortstack{Empty\\Atten.\\(dB) $\uparrow$} \\
    \hline
    DSS~\cite{patterson2022distance} & Fixed & 22.03 & 0.39 & 56.12 \\
    \shortstack[l]{QD-TSE~\cite{shi2025distance}\\+ interval sum} & \shortstack{Local\\+ sum} & 25.99 & 1.12 & 54.10 \\
    Ours & Continuous & \textbf{29.25} & \textbf{4.40} & \textbf{62.01} \\
    \hline
  \end{tabular}
\end{table}

\subsection{Ablation study}
\label{sec:ablation_study}

The ablation in Table~\ref{tab:ablation_study} isolates the contributions of
continuous radius encoding and CRC. The discrete variant uses one-hot radius
codes while all other settings remain unchanged.
Continuous encoding outperforms discrete encoding even without CRC, improving
SI-SDR by 2.36~dB from 24.68~dB to 27.04~dB and reducing RCE from 1.04\% to
0.85\%. 
CRC improves both encoding schemes. With continuous-radius encoding, it
increases SI-SDR by 2.21~dB, SI-SDRi by 1.32~dB, and attenuation by
10.88~dB, while reducing RCE to 0.46\%. Combining continuous-radius encoding
with CRC gives the best value for every metric.
These results confirm that both continuous-radius encoding and CRC learning
contribute to regional extraction performance, with their combination yielding
the strongest results.

\begin{table}[H]
  \caption{Ablation study on radius encoding and CRC.}
  \label{tab:ablation_study}
  \centering
  \small
  \setlength{\tabcolsep}{2.5pt}
  \begin{tabular}{l|c|ccc|c}
    \hline
    \shortstack{Radius\\encoding} &
    CRC &
    \shortstack{Full\\SI-SDR\\(dB) $\uparrow$} &
    \shortstack{Partial\\SI-SDRi\\(dB) $\uparrow$} &
    \shortstack{Empty\\Atten.\\(dB) $\uparrow$} &
    \shortstack{Cross-radius\\RCE (\%) $\downarrow$} \\
    \hline
    Discrete   & $\times$     & 24.68 & 3.04 & 49.68 & 1.04 \\
    Discrete   & $\checkmark$ & 28.75 & 3.15 & 57.70 & 0.81 \\
    Continuous & $\times$     & 27.04 & 3.08 & 51.13 & 0.85 \\
    Continuous & $\checkmark$ & \textbf{29.25} & \textbf{4.40} & \textbf{62.01} & \textbf{0.46} \\
    \hline
  \end{tabular}
\end{table}

\subsection{Continuous-radius analysis}
\label{sec:continuous_radius_analysis}

\begin{figure}[t]
  \centering
  \begin{tabular}{@{}c@{\hspace{0.015\linewidth}}c@{}}
    \includegraphics[height=0.44\linewidth]{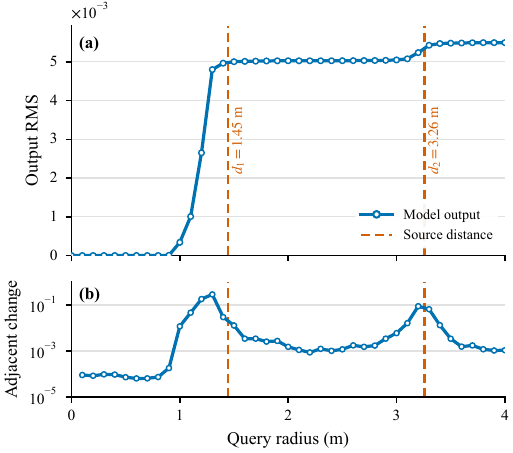} &
    \includegraphics[height=0.44\linewidth]{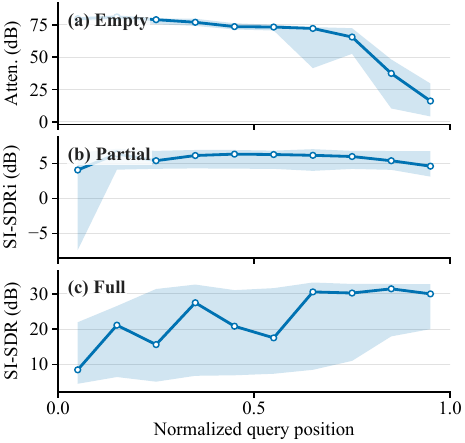} \\[-1mm]
    (a) & (b)
  \end{tabular}
  \caption{Continuous-radius analysis: (a) response of a representative
  mixture and (b) aggregated dense-query performance.}
  \label{fig:continuous_radius}
\end{figure}

Beyond the four shared comparison radii, dense 0.1-m queries examine whether
outputs remain stable within unchanged target regions while preserving
transitions at source boundaries (Fig.~\ref{fig:continuous_radius}).
Fig.~\ref{fig:continuous_radius}(a) shows the normalized output
change for a representative mixture. The output remains stable when the
target-speaker set is unchanged and changes sharply when the radius crosses a
source boundary.
Fig.~\ref{fig:continuous_radius}(b) shows aggregated task-oriented metrics over the
test set. For each mixture with source distances $d_1<d_2$, query positions in
the empty $[0,d_1)$, partial $[d_1,d_2)$, and full $[d_2,R_{\max}]$ regions are
independently normalized to $[0,1]$ and grouped into bins. The curves and
shaded areas denote the median and interquartile range, respectively. The
results show strong empty-region suppression, positive partial-region
SI-SDRi, and preserved full-target SI-SDR across densely sampled radii.

\subsection{Evaluation under challenging conditions}
\label{sec:generalization_analysis}

Finally, Table~\ref{tab:generalization} evaluates generalization to unseen
three-speaker mixtures and robustness to additive noise at 5 and 10~dB SNR.
When the speaker count increases from two to three, SI-SDR 
decreases by 3.08~dB and SI-SDRi by 1.21~dB, while attenuation decreases 
to 59.28~dB and RCE increases from 0.46\% to 1.12\%.
Under additive noise at 10~dB SNR, SI-SDR decreases by 0.20~dB and SI-SDRi
by 0.17~dB, while RCE increases by a factor of five from 0.46\% to 2.31\%. 
At 5~dB SNR, RCE further increases to 3.17\%. 
Under the combined three-speaker, 10-dB condition, the model
achieves 3.10~dB SI-SDRi and 52.67~dB attenuation with an RCE of 4.15\%.

\begin{table}[t]
  \caption{Evaluation under challenging conditions.}
  \label{tab:generalization}
  \centering
  \small
  \setlength{\tabcolsep}{2.5pt}
  \begin{tabular}{l|ccc|c}
    \hline
    Condition &
    \shortstack{Full\\SI-SDR\\(dB) $\uparrow$} &
    \shortstack{Partial\\SI-SDRi\\(dB) $\uparrow$} &
    \shortstack{Empty\\Atten.\\(dB) $\uparrow$} &
    \shortstack{Cross-radius\\RCE (\%) $\downarrow$} \\
    \hline
    2-spk, Clean & 29.25 & 4.40 & 62.01 & 0.46 \\
    2-spk, 10~dB & 29.05 & 4.23 & 61.76 & 2.31 \\
    2-spk, 5~dB  & 28.38 & 3.92 & 60.47 & 3.17 \\
    \hline
    3-spk, Clean & 26.17 & 3.19 & 59.28 & 1.12 \\
    3-spk, 10~dB & 24.98 & 3.10 & 52.67 & 4.15 \\
    \hline
  \end{tabular}
\end{table}

\section{Conclusion}
\label{sec:conclusion}

This work presents a monaural continuous-radius regional speech extraction
method that directly models cumulative regional targets and supports queries
within the trained range using a single model. Ablations show that continuous
radius encoding outperforms one-hot radius codes, while CRC improves both
extraction performance and cross-radius consistency. Dense-query analysis shows
that outputs remain stable when the target-speaker set is unchanged and change
sharply at source boundaries. Evaluations on measured RIRs show that the
proposed method outperforms task-matched fixed-threshold and local-range
baselines and remains effective with more speakers and additive noise.

\bibliographystyle{IEEEbib}
\bibliography{refs}

\begin{thebibliography}{10}

\bibitem{zmolikova2017speakeraware}
Kate\v{r}ina \v{Z}mol\'\i{}kov\'a, Marc Delcroix, Keisuke Kinoshita, Takuya Higuchi, Atsunori Ogawa, and Tomohiro Nakatani,
\newblock ``Speaker-aware neural network based beamformer for speaker extraction in speech mixtures,''
\newblock in {\em Interspeech 2017}. Aug. 2017, pp. 2655--2659, ISCA.

\bibitem{wang2019voicefilter}
Quan Wang, Hannah Muckenhirn, Kevin Wilson, Prashant Sridhar, Zelin Wu, John~R. Hershey, Rif~A. Saurous, Ron~J. Weiss, Ye~Jia, and Ignacio~Lopez Moreno,
\newblock ``{VoiceFilter}: Targeted voice separation by speaker-conditioned spectrogram masking,''
\newblock in {\em Interspeech 2019}. Sept. 2019, pp. 2728--2732, ISCA.

\bibitem{georganti2011speaker}
Eleftheria Georganti, Tobias May, Steven van~de Par, Aki H{\"a}rm{\"a}, and John Mourjopoulos,
\newblock ``Speaker distance detection using a single microphone,''
\newblock {\em IEEE Transactions on Audio, Speech, and Language Processing}, vol. 19, no. 7, pp. 1949--1961, Sept. 2011.

\bibitem{neri2024speaker}
Michael Neri, Archontis Politis, Daniel~A. Krause, Marco Carli, and Tuomas Virtanen,
\newblock ``Speaker distance estimation in enclosures from single-channel audio,''
\newblock {\em IEEE/ACM Transactions on Audio, Speech, and Language Processing}, vol. 32, pp. 2242--2254, 2024.

\bibitem{patterson2022distance}
Katharine Patterson, Kevin Wilson, Scott Wisdom, and John~R. Hershey,
\newblock ``Distance-based sound separation,''
\newblock in {\em Interspeech 2022}. Sept. 2022, pp. 901--905, ISCA.

\bibitem{lin2023focus}
Jiuxin Lin, Peng Wang, Heinrich Dinkel, Jun Chen, Zhiyong Wu, Zhiyong Yan, Yongqing Wang, Junbo Zhang, and Yujun Wang,
\newblock ``Focus on the sound around you: Monaural target speaker extraction via distance and speaker information,''
\newblock in {\em Interspeech 2023}. Aug. 2023, pp. 2488--2492, ISCA.

\bibitem{shi2025distance}
Runwu Shi, Benjamin Yen, and Kazuhiro Nakadai,
\newblock ``Distance based single-channel target speech extraction,''
\newblock in {\em ICASSP 2025 -- 2025 IEEE International Conference on Acoustics, Speech and Signal Processing (ICASSP)}. Apr. 2025, pp. 1--5, IEEE.

\bibitem{shi2025room}
Runwu Shi, Zirui Lin, Benjamin Yen, Jiang Wang, Ragib~Amin Nihal, and Kazuhiro Nakadai,
\newblock ``Single-channel target speech extraction utilizing distance and room clues,''
\newblock in {\em 2025 33rd European Signal Processing Conference (EUSIPCO)}. Sept. 2025, pp. 481--485, IEEE.

\bibitem{gu2024rezero}
Rongzhi Gu and Yi~Luo,
\newblock ``{ReZero}: Region-customizable sound extraction,''
\newblock {\em IEEE/ACM Transactions on Audio, Speech, and Language Processing}, vol. 32, pp. 2576--2589, 2024.

\bibitem{chen2024soundbubbles}
Tuochao Chen, Malek Itani, Sefik~Emre Eskimez, Takuya Yoshioka, and Shyamnath Gollakota,
\newblock ``Hearable devices with sound bubbles,''
\newblock {\em Nature Electronics}, vol. 7, no. 11, pp. 1047--1058, Nov. 2024.

\bibitem{jiang2025regional}
Yiheng Jiang, Haoxu Wang, Yafeng Chen, Gang Qiao, and Biao Tian,
\newblock ``Exploring efficient directional and distance cues for regional speech separation,''
\newblock in {\em Interspeech 2025}. Aug. 2025, pp. 1478--1482, ISCA.

\bibitem{lu2026region}
S.~Lu, X.~Zhou, Y.~Yang, B.~Zhu, and W.~Zhang,
\newblock ``Region-conditioned target speaker extraction via reverberation cues and {Gaussian}-kernel distance encoding,''
\newblock in {\em 2026 34th European Signal Processing Conference (EUSIPCO)}, 2026, pp. 381--385.

\bibitem{wang2023tfgridnet}
Zhong-Qiu Wang, Samuele Cornell, Shukjae Choi, Younglo Lee, Byeong-Yeol Kim, and Shinji Watanabe,
\newblock ``{TF-GridNet}: Integrating full- and sub-band modeling for speech separation,''
\newblock {\em IEEE/ACM Transactions on Audio, Speech, and Language Processing}, vol. 31, pp. 3221--3236, 2023.

\bibitem{perez2018film}
Ethan Perez, Florian Strub, Harm de~Vries, Vincent Dumoulin, and Aaron Courville,
\newblock ``{FiLM}: Visual reasoning with a general conditioning layer,''
\newblock in {\em Proceedings of the AAAI Conference on Artificial Intelligence}, 2018, vol.~32, pp. 3942--3951.

\bibitem{diaz2020gpurir}
David Diaz-Guerra, Antonio Miguel, and Jose~R. Beltran,
\newblock ``{gpuRIR}: A {Python} library for room impulse response simulation with {GPU} acceleration,''
\newblock in {\em 2020 IEEE International Conference on Acoustics, Speech and Signal Processing (ICASSP)}. May 2020, pp. 331--335, IEEE.

\bibitem{dicarlo2021dechorate}
Diego~Di Carlo, Pinchas Tandeitnik, C{\'e}dric Foy, Nancy Bertin, Antoine Deleforge, and Sharon Gannot,
\newblock ``{dEchorate}: A calibrated room impulse response dataset for echo-aware signal processing,''
\newblock {\em EURASIP Journal on Audio, Speech, and Music Processing}, vol. 2021, no. 1, Nov. 2021.

\bibitem{gburrek2024diminishing}
Tobias Gburrek, Adrian Meise, Joerg Schmalenstroeer, and Reinhold Haeb-Umbach,
\newblock ``Diminishing domain mismatch for {DNN}-based acoustic distance estimation via stochastic room reverberation models,''
\newblock in {\em 2024 18th International Workshop on Acoustic Signal Enhancement (IWAENC)}. Sept. 2024, pp. 279--283, IEEE.

\bibitem{bu2017aishell1}
Hui Bu, Jiayu Du, Xingyu Na, Bengu Wu, and Hao Zheng,
\newblock ``{AISHELL-1}: An open-source {Mandarin} speech corpus and a speech recognition baseline,''
\newblock in {\em 2017 20th Conference of the Oriental Chapter of the International Coordinating Committee on Speech Databases and Speech I/O Systems and Assessment (O-COCOSDA)}. Nov. 2017, pp. 1--5, IEEE.

\bibitem{du2018aishell2}
Jiayu Du, Xingyu Na, Xuechen Liu, and Hui Bu,
\newblock ``{AISHELL-2}: Transforming {Mandarin} {ASR} research into industrial scale,''
\newblock {\em arXiv preprint arXiv:1808.10583}, 2018.

\bibitem{wang2015thchs30}
Dong Wang and Xuewei Zhang,
\newblock ``{THCHS-30}: A free {Chinese} speech corpus,''
\newblock {\em arXiv preprint arXiv:1512.01882}, 2015.

\bibitem{yamagishi2019vctk}
Junichi Yamagishi, Christophe Veaux, and Kirsten MacDonald,
\newblock ``{CSTR VCTK} corpus: English multi-speaker corpus for {CSTR} voice cloning toolkit (version 0.92),'' 2019,
\newblock Dataset, doi: 10.7488/ds/2645.

\bibitem{dubey2024dns}
Harishchandra Dubey, Ashkan Aazami, Vishak Gopal, Babak Naderi, Sebastian Braun, Ross Cutler, Alex Ju, Mehdi Zohourian, Min Tang, Mehrsa Golestaneh, and Robert Aichner,
\newblock ``{ICASSP} 2023 deep noise suppression challenge,''
\newblock {\em IEEE Open Journal of Signal Processing}, vol. 5, pp. 725--737, 2024.

\bibitem{leroux2019sdr}
Jonathan~Le Roux, Scott Wisdom, Hakan Erdogan, and John~R. Hershey,
\newblock ``{SDR} -- half-baked or well done?,''
\newblock in {\em 2019 IEEE International Conference on Acoustics, Speech and Signal Processing (ICASSP)}. May 2019, pp. 626--630, IEEE.

\end{thebibliography}

\end{document}